\documentstyle[aps,pra,eqsecnum,twocolumn,epsfig,tabularx,array]{revtex}

\begin{document}
\draft
\wideabs{
\title{Optimum unambiguous discrimination between linearly-independent
non-orthogonal quantum states and its optical realization}
\author{Yuqing Sun$^{1}$, Mark Hillery$^{1}$, and J\'anos A. 
Bergou$^{1,2}$}
\address{$^1$Department of Physics, Hunter College, City
    University of New York, 695 Park Avenue, New York, NY 10021,
    USA \\
    $^{2}$Institute of Physics, Janus Pannonius University,
    H-7624 P\'{e}cs, Ifj\'{u}s\'{a}g \'{u}tja 6, Hungary}
\date{\today}
\maketitle
\begin{abstract}
Unambiguously distinguishing between nonorthogonal but linearly
independent quantum states is a challenging problem in
quantum information processing. In principle, the problem can be
solved by mapping the set of nonorthogonal quantum states onto a 
set of orthogonal ones, which then can be distinguished without 
error. Such nonunitary transformations can be performed 
conditionally on quantum systems; a unitary transformation is
carried out on a larger system of which the system of interest
is a subsytem, a measurement is performed, and if the proper
result is obtained, the desired nonunitary transformation will
have been performed on the subsystem.  We show how to construct 
generalized interferometers (multiports), which when combined 
with measurements on some of the output ports, implement
nonunitary transformations of this type. The input states are
single-photon states in which the photon is divided among several
modes.  A number of explicit examples of distinguishing among
three nonorthogonal states are discussed, and the networks that 
optimally distinguish among these states are presented.
\end{abstract}
}

\section{Introduction}

The time evolution of a closed quantum system is unitary, 
hence scalar products (angles between two quantum states) are 
conserved. However, when measurements are made on the system, 
it is possible to perform
prescribed non-unitary operations with a certain probability
of success. In particular, it is possible to alter the
value of scalar products and, hence, the angle between state
vectors. Such operations can be used to unambiguously
discriminate among non-orthogonal quantum states. A set of 
non-orthogonal states is mapped onto a set of orthogonal ones,
and the orthogonal states can be distinguished without error.
According to the quantum theory of measurement, such a non-unitary
transformation will always have a certain probability of failure, 
which, for the discrimination
of non-orthogonal states, corresponds to the probability that we 
obtain inconclusive answers. Our aim here is to find the 
optimum solution that minimizes the average probability of 
failure.

Considerable work has been done on this problem.  The simplest 
case, distinguishing two nonorthogonal states was first considered
by Ivanovic \cite{ivanovic}, and then subsequently by Dieks
\cite{dieks} and Peres \cite{peres2}.  These authors found
the optimal solution when the two states are being selected
from an ensemble in which they are equally likely.  The optimal
solution for the situation in which the states have different 
weights was found by Jaeger and Shimony \cite{jaeger}. One
can also consider what happens if the discrimination is not
completely unambiguous, i.\ e.\ if it is possible for errors
to occur, and this was done by Chefles and Barnett 
\cite{chefles1}. The
case of three states was examined by Peres and Terno 
\cite{peres3}.  The general $N$-state problem has been studied by
Chefles \cite{chefles2}, by Chefles and Barnett \cite{chefles3},
and by Duan and Guo \cite{duan}.  Chefles and Barnett employed
the POVM formalism and specifically solved the case in which
the probability of the procedure succeeding is the same for
each of the states.  Duan and Guo considered general unitary
transformations and measurements on a Hilbert space containing
the states to be distinguished and an ancilla,
which would allow one to discriminate among $N$ states, and
derived matrix inequalities which must be satisfied for the
desired transformations to exist.

For the experimental realization of quantum information processing,
one must choose a physical system to represent a qubit.  Some 
possibilities which have been
used are energy levels of ions, the orientation of a nuclear
spin, and the presence or absence of a photon in a cavity
\cite{cirac}--\cite{haroche}.
Another possibility, the so-called dual-rail
representation of a qubit, was proposed by Milburn 
\cite{milburn}, and later
by Chuang and Yamamoto \cite{chuang1,chuang2}.  A photon is
split between two modes which represent $0$ and $1$. When the
elementary carriers of the information are more than two dimensional
objects (qutrits,$\ldots$, qunits in general, for $n$
dimensions), one needs a more general representation. Here we will
show that single photon states can be used to represent general
non-orthogonal states in n dimensions, and how this representation
can be used for state discrimination. A photon
is now divided among $n$ modes which represent 
the numbers $0,1,\ldots,n-1$.
The method is a straightforward generalization of the dual rail
representation of a qubit for more than two dimensions, and can be
called the multiple rail representation of a qunit. An optical
multiport, a kind of a generalized interferometer with more than two
inputs and outputs, together with measurements made by the photon
detectors placed at some or all of the output ports, can 
conditionally realize the desired non-unitary transformations 
of the initial, non-orthogonal single-photon states into
orthogonal states. Our previous paper \cite{bergou} which proposed 
an optical realization to optimally discriminate between two 
non-orthogonal
states is a special case of the method presented here.

Optical experiments to distinguish between two quantum states 
have already been carried out, first by Huttner, \emph{et}.\ \emph{al}.\ 
\cite{huttner} and, more recently, by Clarke, \emph{et}.\ \emph{al}. 
\cite{clarke}.  Both of these used the polarization states
of photons to represent qubits.

This paper is divided into six sections. In Sec. II, we 
present a method for calculating the optimum probabilities 
of unambiguous discrimination between linearly-independent,
non-orthogonal states. In Sec. III, the general properties
of a quantum system, which realizes the optimum non-unitary
transformation, are found by assuming that the optimum probabilities
are known. In Sec. IV, we will show how an optical multiport, 
which is designed to perform a particular 
unitary transformation, together with measurements at its output
ports can realize non-unitary transformations of non-orthogonal
input states represented by single photon states. Reck {\it{et al.}}
\cite{reck} gave a method to decompose any multiport into a series
of beam splitters, phase shifters, and mirrors, and we will use 
this method to
construct the desired multiport. The general method is then
illustrated by applying it to qutrits
in Sec. V. Examples are presented for the realization
of transformations that convert three specific non-orthogonal
states to orthogonal ones with a maximum probability of success. A
brief discussion and conclusions are given in Sec. VI.

\section{Optimal probabilities for unambiguous discrimination 
among non-orthogonal quantum states}

Suppose we are given a quantum system prepared in the state 
$|\psi\rangle$, which is guaranteed to be a member of the set of non-orthogonal states $\left\{
|\psi_{1}\rangle,|\psi_{2}\rangle,\ldots,|\psi_{n}\rangle\right\}$,
but we do not know which one.  We want to find a procedure which 
will tell us which member of the set we were given.  The procedure
may fail to give us any information about the state, and if it 
fails, it must let us know that it has, but if it succeeds, it 
should never give us a wrong answer.  We shall refer to such a
procedure as state discrimination without error.  Note that this
procedure has $n+1$ outcomes; it either tells us which state
we were given, or it tells us that it failed (inconclusive
outcome).

In order to achieve error-free discrimination, Chefles has shown,
in a very clear analysis of the problem,
that the set $|\psi_{1}\rangle,|\psi_{2}\rangle,\ldots,|\psi_{n}
\rangle$ must be linearly independent \cite{chefles2}.  If the 
states are not orthogonal (which we shall assume), they cannot be discriminated perfectly. 
That means that if we are given $|\psi_{i}\rangle$, we will have 
some probability $p_{i}$ to distinguish it successfully and, 
correspondingly, some
failure probability $q_{i}= 1-p_{i}$ to obtain an inconclusive
answer. Denote by $\mathcal{H}$ the Hilbert space spanned by
the initial states $\left\{ |\psi_{1}\rangle,
|\psi_{2}\rangle,\ldots,|\psi_{n}\rangle \right\}$. Since there 
is a chance to get an inconclusive answer, the number of outcomes 
of this process is larger than the dimension of $\mathcal{H}$, 
hence this process is a ``generalized measurement'' which can be 
represented by a set of operators which
form a resolution of the identity  \cite{chefles2},
\begin{equation}
\label{a2}
{\hat A}^{\dagger}_{I}{\hat A}_{I}+\sum_{i}{\hat
A}^{\dagger}_{i}{\hat
A}_{i}={\hat {\bf 1}} ,
\end{equation}
where ${\hat A}_i$ is the operator that corresponds to the 
outcome $|\psi_{i}\rangle$, and ${\hat A}_I$ is the operator that
corresponds to the inconclusive outcome.  In more detail, if
$\rho$ is the density matrix of our given state, then the
probability of obtaining the $k$th outcome, where $k$ can be
$1,\ldots n$ or $I$, is $p_{k}=Tr(\rho A^{\dagger}_{k}A_{k})$ and
if the outcome is $k$, then the resulting density matrix is
$A_{k}\rho A_{k}^{\dagger}/p_{k}$.  The requirement that the
discrimination be error free implies that
\begin{equation}
\langle\psi_{i}|A_{k}^{\dagger}A_{k}|\psi_{i}\rangle =p_{i}
\delta_{ik} .
\end{equation}
From this, by an application of the Schwarz inequality, it follows 
that
\begin{equation}
\label{a5}
\langle \psi_k |{\hat A}_i^{\dagger} {\hat A}_i| \psi_j \rangle
= p_i \delta_{ji} \delta_{jk},
\end{equation}

If we denote by $\eta_i$ the {\em a priori} probability that the
system was
prepared in the state $|\psi_{i}\rangle$, the average probabilities
of success and of failure to distinguish the states $|\psi_{i}\rangle$ 
are, respectively,
\begin{eqnarray}
\label{Psf}
P &=& \sum_{i}{\eta}_{i}p_{i} , \nonumber \\
Q &=& \sum_{i}{\eta}_{i}q_{i} .
\end{eqnarray}
Our objective is to find the set of $\left\{ p_i \right\}$ that
maximizes the probability of success, $P$, and the set of 
operators $A_{k}$ that realize the corresponding generalized
measurement.

Define the failure state, $|\phi_{i}\rangle$ as
\begin{equation}
\label{a3}
|\phi_{i}\rangle = {\hat A}_I |\psi_{i}\rangle .
\end{equation}
This is the state of the system if the input state was
$|\psi_{i}\rangle$ and the outcome was inconclusive.
Chefles \cite{chefles2} showed that the states
$\left\{|\phi_{i}\rangle\right\}$ are linearly dependent when
$P$ is a maximum.  The interpretation of this result is the 
following.  Because only 
linearly independent states can be discriminated without error,
the operator corresponding to the inconclusive outcome maps the 
set of linearly dependent states $\left\{|\psi_{i}\rangle\right\}$ 
onto a linearly-dependent set, which then cannot be unambiguously 
discriminated by any further process.  As we shall see, however,
this does not mean that some information cannot be extracted
from an inconclusive result.

Now consider the inner product $\langle \phi_k|\phi_j \rangle$,
and define the matrix $C$ by $C_{ij}=\langle \phi_k|\phi_j 
\rangle$.  Using equations
\ (\ref{a2}), \ (\ref{a3}) and \ (\ref{a5}), we find
\begin{equation}
\label{a6}
\langle \phi_k|\phi_j \rangle = \langle \psi_k|\psi_j \rangle
- p_j \delta_{jk}.
\end{equation}
The matrix $C$ is positive-semidefinite. This can be seen by
noting that for any n-dimensional vector, whose components we shall
denote by $b_{i}$, where $i=1,\ldots n$,
\begin{equation}
\sum_{i,j=1}^{n}b_{i}^{\ast}C_{ij}b_{j} =\|\sum_{i=1}^{n}b_{i}
|\phi_{i}\rangle\|^{2}\ge 0.
\end{equation}
When the $p_{i}$ are equal to their optimal values, i.\ e.\ the values 
which maximize $P$,  the linear dependence of the $|\phi_{i}\rangle$
implies that
\begin{equation}
\label{a8}
\det (C) =0 ,
\end{equation} 
so that $C$ has at least one zero eigenvalue when $P$ is a maximum.

In conclusion, the optimum probabilities $p_{i}$ can be found by
maximizing $P$ subject to the following constraints:
\\

i) det($C$) = 0,

ii) $C$ is non-negative, or, equivalently, all of the principal
minors of $C$ are non-negative.
 \\

When we consider the case of just two non-orthogonal states, the
above result 
immediately gives the following relationship between any two
failure probabilities:
\begin{equation}
\label{a12}
q_1 \cdot q_2=|\langle \psi_1|\psi_2 \rangle |^{2}.
\end{equation}
This is the same result that was obtained in our previous paper
\cite{bergou}. In particular, when the two
states have equal {\em a priori} probabilities,
$\eta_1 = \eta_2 = \frac{1}{2}$, we found the
maximum probability of success to be,
\begin{equation}
\label{a13}
P = 1-|\langle \psi_1|\psi_2 \rangle |.
\end{equation}
This is the well known Ivanovic-Dieks-Peres (IDP) limit
\cite{ivanovic}--\cite{peres2}.

\section{Realization of generalized measurement}

Once we know the set of optimum
discrimination probabilities $\left\{ p_i \right\}$, we would
like to find a realizable experimental procedure to achieve it.
We shall do this first abstractly, and then show how it can be
realized by linear optical elements.  Let us first summarize 
the procedure, and subsequently fill in the details.  We begin 
with a total Hilbert space $\mathcal{K}$, which is the direct sum 
of two subspaces, $\mathcal{K}=\mathcal{H}\oplus\mathcal{A}$.  The 
space $\mathcal{H}$ is an $n$-dimensional space that contains the 
vectors $|\psi_{i}\rangle$, and $\mathcal{A}$ is the space that
will contain the failure vectors $|\phi_{i}\rangle$.  We shall denote the
dimension of $\mathcal{A}$ by $m$.
The input state of the system is one of the vectors $|\psi_{i}\rangle$,
which is now a vector in the subspace $\mathcal{H}$ of the total
space $\mathcal{K}$.  A unitary transformation, $U$, which acts in
the entire space $\mathcal{K}$ is now applied to the input vector,
resulting in the state $|\psi_i^{\mathcal K} \rangle_{out}$. A 
measurement is performed on the part of 
$|\psi_i^{\mathcal K} \rangle_{out}$ 
in $\mathcal{A}$, and, if the proper result is obtained, the 
vector $|\psi_i^{\mathcal K} \rangle_{out}$ is projected onto the
vector $|e_i^{\mathcal H} \rangle$, which lies in the subspace
$\mathcal{H}$.  The probability of this occuring is $p_i$.  The 
vectors $\{ |e_i^{\mathcal H} \rangle, i=1,\ldots n\}$ are 
orthonormal and can be distinguished perfectly.  The effect of
the unitary transformation on an extended space and the measurement is
to map a set of non-orthogonal vectors onto a set of orthogonal ones.

We now need to specify $U$ and the measurement, and let us discuss
the latter first.  The measurement has two outcomes, 
one of them corresponding to the operator, $P_{\mathcal{H}}$,
which projects onto the subspace $\mathcal{H}$, and the other to the
operator $P_{\mathcal{A}}=I-P_{\mathcal{H}}$, which projects onto the
subspace $\mathcal{A}$.  The first outcome
corresponds to the successful transformation of $|\psi_i^{\mathcal K} 
\rangle_{out}$ into $|e_i^{\mathcal H} \rangle$, and its 
probability of occurence is $p_{i}$.  This implies that
\begin{equation}
\label{b2}
|\psi_i^{\mathcal K} \rangle_{out} = \sqrt {p_i} |e_i^{\mathcal H}
\rangle
 + 
|\phi_i^{\mathcal A} \rangle ,
\end{equation}
where $|\phi_i^{\mathcal A} \rangle$ is a failure state, and we
have added a superscript $\mathcal{A}$ to denote the fact that
it is in the subspace $\mathcal{A}$.  The other outcome 
corresponds to obtaining an inconclusive answer and transforms 
$|\psi_i^{\mathcal K} \rangle_{out}$ into $|\phi_i^{\mathcal A} 
\rangle$.

Eq. (\ref{b2}) and the fact that $U$ is unitary implies that 
\begin{equation} 
\label{b21} 
\langle \phi_k^{\mathcal A}|\phi_j^{\mathcal A}\rangle
=  \langle \psi_k^{\mathcal H}|\psi_j^{\mathcal H} \rangle
- p_j \delta_{jk},
\end{equation}
which is just Eq. (\ref{a6}).  If (and only if) the matrix $C$ is
non-negative, we can always find vectors $|\phi_i^{\mathcal A} 
\rangle$ that satisfy this equation.  This follows from the
fact that a non-negative matrix can be written as the product
of a matrix and its adjoint, in particular, we can express
$C$ as
\begin{equation}
C=A^{\dagger}A ,
\end{equation}
for some matrix $A$.  If we define $|\phi_{j}\rangle =A|j\rangle$,
where $|j\rangle$ is the vector whose $j$th component is one and
all of whose other components are zero, then we have that
$C_{jk}=\langle\phi_{j}|\phi_{k}\rangle$ .  Once we have found
these vectors and specified the vectors $|e_{i}^{\mathcal{H}}
\rangle$, then the operator $U$ can be found by means of Eq.
(\ref{b2}).  These conditions may not completely determine $U$;
if they do not, then there is freedom in choosing it.  This will 
be the case if the dimension of $\mathcal{A}$ is greater than 
one.  $U$ maps vectors in $\mathcal{A}$ to vectors in the 
subspace, $\mathcal{S}$, of $\mathcal{K}$ that consists of the
vectors that are orthogonal to all of the vectors 
$|\psi_i^{\mathcal K} \rangle_{out}$.  The dimension of 
$\mathcal{S}$ is $m$.  The freedom in choosing $U$ comes from the
fact that Eq. (\ref{b2}) does not specify how $\mathcal{A}$ is
mapped into $\mathcal{S}$.  If both are one-dimensional, then the
mapping is determined (up to an overall phase), but if their
dimension is greater than two it is not.

Once the measurement and the
operator $U$ have been specified, our realization of the 
generalized measurement is completely determined.  The next task 
is to find a physical system with which to implement it.

\section{Optical realization of non-unitary transformation}

We now want to propose an experimental procedure to achieve our
non-unitary transformation by using optical devices. We shall 
show how this can be accomplished by using a single-photon 
representation of the states $|\psi_{i}\rangle$ and an optical 
multiport together with photodetectors at the output ports to
carry out the desired non-unitary transformation.

Our Hilbert space will consist of a single photon, which is
divided among $n+m$ modes.   
The modes themselves could be
distinguished by having different wave vectors or they might be
modes of different optical fibers.
A basis for this space consists
of the single photon states $\{ a_{j}^{\dagger}|0\rangle |
j =1,\ldots n+m\}$, where $|0\rangle$ is the vacuum state
and $a_{j}^{\dagger}$ is the creation operator for the 
$j$th mode.  The states $\{ a_{j}^{\dagger}|0\rangle |
j =1,\ldots n\}$ form a basis for the space $\mathcal{H}$,
and the states $\{ a_{j}^{\dagger}|0\rangle |
j =n+1,\ldots n+m\}$ form a basis for the space $\mathcal{A}$.
The initial states $|\psi_i \rangle$ can be
represented as single photon states in $\mathcal{H}$, which 
can be written as
\begin{equation}
\label{c1}
|\psi_i \rangle =\sum_{j=1}^{n}d_{i j}
|e_{j}^{\mathcal H} \rangle
=\sum_{j=1}^{n}d_{i j} \hat{a}_{j}^{\dagger} |0\rangle,
\end{equation}
where we have chosen the states $|e_{j}^{\mathcal H} \rangle$
to be $|e_{j}^{\mathcal H} \rangle = \hat{a}_{j}^{\dagger}
|0^{\mathcal H} \rangle$.

An optical $2N$-port is a lossless linear device
with $N$ input ports and $N$ output
ports. Its action on the input states can be described by a 
unitary operator,
$U_{2N}$, and physically it consists of an arrangement of beam 
splitters, phase shifters, and mirrors. Choosing
$N=n+m$, we send the single photon state $|\psi_i\rangle$ 
into the first $n$ input ports, which correspond to
$\mathcal{H}$, and the vacuum into the remaining $m$ input ports,
which correspond to $\mathcal{A}$. Photodetectors are placed 
at the last $m$ output ports (the ones corresponding to 
$\mathcal{A}$), and if there is no
photon detected, the desired non-unitary transformation will
have been carried out.  In particular, with $|\psi_i^{\mathcal K} 
\rangle_{out}=U_{2N}|\psi_{i}\rangle$, 
where $|\psi_i^{\mathcal K} 
\rangle_{out}$ is given by Eq. (\ref{b2}), the action of the
measurement, if successful, is to project the
output state onto $|e_{i}^{\mathcal H} \rangle$, and
the probability to achieve this is $p_i$.

If we denote the annihilation operators corresponding to the input
modes of the $2N$-port by 
$a_{j \bf{in}}$, $j=1,2,\ldots,N$, then the output operators are
given by
\begin{equation}
\label{trans}
a_{j\bf{out}}=U^{-1}_{2N}a_{j\bf{in}}U_{2N} 
=\sum_{k=1}^{N}M_{jk}a_{k\bf{in}},
\end{equation}
where $M_{jk}$ are the elements of an $N\times N$ unitary 
matrix $M$.  In the 
Schr\"{o}dinger picture, the $in$ and $out$ states are
related by
\begin{equation}
|\psi^{\mathcal K} \rangle_{out} = U_{2N}|\psi^{\mathcal K} 
\rangle_{in} .
\end{equation}
In general, for an $in$ state that contains a single photon
\begin{equation}
\label{c3}
|\psi^{\mathcal K} \rangle_{in}
=\sum_{j=1}^{N} c_j a_j^{\dagger}|0\rangle ,
\end{equation}
where $\sum_{j=1}^{N}|c_{j}|^{2}=1$, the $out$ state is given by
\begin{eqnarray}
\label{out1}
|\psi^{\mathcal K} \rangle_{out}
 & =  & U_{2N} |\psi^{\mathcal K} \rangle_{in} \nonumber \\
 & =&  U_{2N}\sum_{j=1}^{N}c_{j}a_{j\bf{in}}^{\dagger}
U^{-1}_{2N}|0\rangle \nonumber \\
 & =&  \sum_{j,k=1}^{N} c_{j}M_{jk}^{T}
a^{\dagger}_{k\bf{in}}|0\rangle .
%\end{align}
\end{eqnarray}
Note that we have made use of the fact 
that the vacuum is invariant under the 
transformation, $U_{2N}$.  This implies that the matrix elements
$M_{il}$ is the same as the matrix element of $U_{2N}$ between the
single-particle states $|i\rangle = a^{\dagger}_{i\bf{in}}|0\rangle$ and 
$|l\rangle = a^{\dagger}_{l\bf{in}}|0\rangle$.  Choosing $c_{j}=
\delta_{jl}$ in the above equation and then taking the inner 
product of the result with $|i\rangle$, we find that
\begin{equation}
\langle i|U_{2N}|l\rangle = M_{il} .
\end{equation}
The desired matrix $M$ can
be found from Eq. (\ref{b2}), and our next task is to
decompose it in such a way that it corresponds to a 
linear optical network.

This problem has been solved by M. Reck {\it{et al.}} \cite{reck},
and we shall summarize their method.  They gave an algorithmic 
procedure to factorize any 
$N \times N$ unitary matrix into a product of two-dimensional
$U(2)$ transformations, and it is this procedure that we shall 
adopt here to construct our $2N \!$-port, which is characterized
by the matrix $M$ of equation\ (\ref{trans}).

It is well known that a lossless beam splitter and a phase
shifter with appropriate parameters can implement any $U(2)$
transformation; a beam splitter with a
phase shifter at one output port transforms the input operators 
into output operators as
\begin{equation}
\label{c8}
\left( \begin{array}{c}
a_1 \\ a_2
\end{array} \right)_{out} = \left(
\begin{array}{cc}
e^{i \phi} \sin \omega & e^{i \phi} \cos \omega \\
\cos \omega & - \sin \omega
\end{array} \right) \left(
\begin{array}{c}
a_1 \\ a_2
\end{array}\right)_{in}
\end{equation}
where $a_1$, $a_2$ are the annihilation operators of modes 1 
and 2 respectively.  In their paper, Reck, {\it{et al.}} considered the
use of a Mach-Zehnder interferometer to simulate the effect of a beam splitter
that does not split the incoming beam equally, in which case $\omega$ describes 
the reflectivity and
transmittance of the effective beam splitter with $\sqrt{R} = \sin \omega$,
$\sqrt{T} = \cos \omega$, and $\phi$ describes the effect of
the phase shifter.  If the matrix describes an actual beam splitter, then
$\sqrt{R} = \cos \omega$, and $\sqrt{T} = \sin \omega$.
Any $N\times N$ unitary matrix $U(N)$ can 
be reduced to an $(N-1)\times (N-1)$ unitary matrix,
$U(N-1)$, by multiplying from the right by  a succession
of two-dimensional unitary matrices
\begin{equation}
\label{c9}
U(N)\cdot R(1) = \left(
\begin{array}{cc}
e^{i \alpha_1} & 0 \\ 0 & U(N-1)
\end{array} \right) .
\end{equation}
Here $R(1)=T_{1,2} \cdot T_{1,3} \cdots T_{1,N}$, and $T_{p,q}$
is defined as an $N \!$-dimensional identity matrix with elements
$I_{pp}$, $I_{pq}$, $I_{qp}$, $I_{qq}$ replaced by the 
corresponding elements of a $U(2)$ matrix. It performs a unitary 
transformation on a two-dimensional subspace of the full $N$
dimensional space, and can be implemented by attaching
a beam splitter and a phase shifter to ports 
$p$ and $q$.

We can repeat the above transformation, decreasing the
dimension of the remaining unitary matrix by one at each step.
Applying this procedure to the matrix $M$ of 
equation\ (\ref{trans}), we have that
\begin{eqnarray}
\label{c10}
M &\cdot& R(1) \cdot R(2) \cdots R(n+m) \nonumber \\
&=& \left(
\begin{array}{cccc}
e^{i \alpha_1} & 0 &  & 0 \\ 0 & e^{i \alpha_2} & & \\ 
0& 0 & \ddots & \\  0 & 0 & &  e^{i\alpha_{n+m}}
\end{array} \right) .
\end{eqnarray}
Denoting by  $D( \alpha_1,\alpha_2\,\ldots \alpha_n )$ the
diagonal matrix
\begin{equation}
\label{c12}
D= \left(
\begin{array}{cccc}
e^{-i \alpha_1} & & 0 & \\  & e^{-i \alpha_2} & & \\ 0 & & 
\ddots & \\ & & & e^{-i \alpha_{n+m}}
\end{array} \right) ,
\end{equation}
we have
\begin{equation}
\label{c13}
M\cdot R(1) \cdot R(2) \cdots R(n+m-1) \cdot D = {\bf 1},
\end{equation}
i.e., 
\begin{equation}
\label{c14}
M=D^{-1}\cdot R(n+m-1)^{-1} \cdots R(1)^{-1} .
\end{equation}

Since the product of matrices is equivalent to setting up
experimental 
devices in sequence, Eq.\ (\ref{c14}) implies that to get $M$,
the actual experimental setup is made of a series of $U(2)$ 
blocks to achieve $R(n+m)^{-1} \cdots R(1)^{-1}$, and $n+m$ 
appropriate phase shifters attached to the output ports 
to produce $D^{-1}$. 
Figure 1 gives a picture of the practical implementation 
of $M$.  

It is possible to save some steps by modifying this procedure.  As
was mentioned earlier, the matrix $M$ is not always completely 
determined.  In particular, there is freedom in choosing the matrix
elements $M_{jk}$ for $k>n$.  Let us now see what happens if we
apply the procedure of Reck, et al.\ to the transpose of $M$, 
$M^{T}$, instead of $M$ itself.  It is now the matrix elements
$(M^{T})_{jk}$, for $j>n$ that are not completely determined, and 
we shall leave them that way for now.  The matrixes making up 
$R(1)$ are chosen to make all of the elements, except the first,
of the the first row of $M^{T}R(1)$ zero.  In finding each of the
matrixes $T_{1,q}$, we only need to use the matrix elements
that are completely determined (this is not true if we start with
$M$ instead of $M^{T}$).  Now if the first row of $M^{T}R(1)$ is
zero except for the first element, then by unitarity, the first
column is also zero, except for its first element.  Continuing 
in this way we have that
\begin{eqnarray}
\label{c15}
M^{T} &\cdot& R(1) \cdot R(2) \cdots R(n) \nonumber \\
&=& \left(
\begin{array}{cccc}
e^{i \alpha_1} & 0 &  & 0 \\ 0 & e^{i \alpha_2} & & \\ 
0& 0 & \ddots & \\  0 & 0 & &  M_{m}
\end{array} \right) ,
\end{eqnarray}
where $M_{m}$ is an $m\times m$ unitary matrix that contains
the information about the matrix elements in $M$ that are
not completely specified.  At this point, we can choose $M_{m}$
to be any unitary matrix, and the simplest choice is the $m
\times m$ identity matrix, $I_{m}$.  Defining
\begin{eqnarray}
\label{c16}
D'&=& \left(
\begin{array}{cccc}
e^{-i \alpha_1} & 0 &  & 0 \\ 0 & e^{-i \alpha_2} & & \\ 
0& 0 & \ddots & \\  0 & 0 & &  I_{m}
\end{array} \right) ,
\end{eqnarray}
we have that
\begin{equation}
\label{mfac}
M=[R(1)R(2)\cdots R(n)D']^{\ast} .
\end{equation}

\section{Application to three states and examples}

In this section we first apply the above considerations to 
the problem of realizing optimal discrimination among three 
non-orthogonal but linearly-independent quantum states, in 
general. Then we illustrate the method on specific examples. 
For simplicity, we assume that
the {\em a priori} probabilities are all equal,
$\eta_{1}=\eta_{2}=\eta_{3}=1/3$.

From Eq.\ (\ref{Psf}), the probability of failure is
\begin{equation}
\label{e1}
Q=\frac{1}{3} \sum_{i=1}^{3} q_{i}.
\end{equation}
The requirement of the linear dependence of the $|\phi_{i}
\rangle$ vectors $(i=1,2,3)$ leads to the constraint given 
by Eq.\ (\ref{a8}). For
the case of three vectors it can be written as
\begin{eqnarray}
\label{e2}
%\begin{split}
\Delta &=& det(C)  \nonumber \\
       &=& q_1 q_2 q_3 - q_1 |O_{23}|^{2}- q_2 |O_{13}|^{2}
           - q_3 |O_{12}|^{2} \nonumber \\
       & & \quad + O_{12} O_{23} O_{13}^{\ast}
           + O_{12}^{\ast} O_{23}^{\ast} O_{13} \nonumber \\
       &=& 0 ,
%\end{split}
\end{eqnarray}
where $O_{ij}=\langle \psi_i|\psi_j \rangle$.

Employing the Lagrange multiplier method, we wish to 
minimize the quantity
\begin{equation}
\label{e3}
Q^{\prime}=\frac{1}{3} \sum_{i} q_{i} +\lambda \Delta ,
\end{equation}
which immediately leads to the conditions
\begin{eqnarray}
\label{e4}
%\begin{cases}
\frac{\partial Q^{\prime}}{\partial q_1} &=&
\frac{1}{3}+ \lambda \Delta_{23} = 0, \nonumber \\
\frac{\partial Q^{\prime}}{\partial q_2} &=&
\frac{1}{3}+ \lambda \Delta_{13} = 0, \nonumber \\
\frac{\partial Q^{\prime}}{\partial q_3} &=&
\frac{1}{3}+ \lambda \Delta_{12} = 0 ,
%\end{cases}
\end{eqnarray}
where $\lambda$ is a Lagrange multiplier,
and $\Delta_{12}$, $\Delta_{13}$, $\Delta_{23}$
are subdeterminants of $C$,
$\Delta_{12}=q_1 q_2-|O_{12}|^{2}$, etc.
Equation\ (\ref{e4}) implies that
\begin{equation}
\label{e5}
\Delta_{12}=\Delta_{13}=\Delta_{23}=- \frac{1}{3 \lambda}.
\end{equation}
This means that all three subdeterminants are equal. 
Let $\delta = -\frac{1}{3 \lambda}$ denote this common value 
and recall that all subdeterminants of $C$ must be 
non-negative, so that $\delta \geq 0$.

From Eq.\ (\ref{e5}) we can solve for the $q_i$'s, yielding
\begin{eqnarray}
%\begin{cases}
\label{e6}
q_1  &=& \sqrt{\frac{(|O_{12}|^{2}+\delta)(|O_{13}|^{2}+\delta)}
{(|O_{23}|^{2}+\delta)}} , \nonumber \\
q_2  &=& \sqrt{\frac{(|O_{12}|^{2}+\delta)(|O_{23}|^{2}+\delta)}
{(|O_{13}|^{2}+\delta)}} , \nonumber \\
q_3  &=& \sqrt{\frac{(|O_{13}|^{2}+\delta)(|O_{23}|^{2}+\delta)}
{(|O_{12}|^{2}+\delta)}} .
%\end{cases}
\end{eqnarray}
Finally, we can substitute Eq.\ (\ref{e6}) into
Eq.\ (\ref{e2}) to solve for $\delta$ and then use the above
equations to find the corresponding $q_i$ values.  When
we solve for $\delta$, there are often a number of different
solutions.  However, we need only consider solutions that are
greater than or equal to zero, and which give values of $q_{i}$
that are between $0$ and $1$.  If there are several solutions
that satisfy these conditions, we must determine which one
gives the actual minimum.  If there are none, then we must
examine the boundary of the allowed region to find the 
minimum.  The point $(q_{1}, q_{2}, q_{3})$ lies inside or
on the surface of a unit cube one whose vertices lie on the
points $(j,k,l)$, where $j,k,l = 0$ or $1$.  If the 
Lagrange multiplier approach does not yield a valid solution
the minimum of $Q$ subject to the constraint $\Delta = 0$
must lie on the surface of the cube.

Note that if the overlaps are real and positive, a situation we shall
consider shortly, then $\delta =0$ is always a solution of Eq.\ (\ref{e2}). 
In this case, if all the corresponding $q_i$ for $\delta=0$ are between $0$
and $1$, then
this set of $\left\{ q_i \right\}$ is a possible solution
to our problem, i.e.\ a minimum of $Q$ that satisfies
$\Delta = 0$.  If it is, in fact the solution, we see that $\Delta_{12}=\Delta_{13}=\Delta_{23}=0$,
which implies that each possible pair of the states $|\phi_{i}\rangle$,
$i=1,2,3$ is linearly dependent, so that all three states
$\phi_i$ are in a line, i.e.\ the dimensionality of the auxiliary
Hilbert space ${\mathcal A}$ is one.  If the solution to the
problem is one for which $\delta >0$, no pair of failure states 
is linearly dependent. However, the three failure states together 
are linearly dependent, so that in this case the dimensionality 
of the auxiliary Hilbert space ${\mathcal A}$ is two.

Next, we shall consider specific examples involving three
non-orthogonal but linearly independent state vectors, to 
illustrate the general considerations of the previous sections. 
In particular we want to determine explicitly the parameters and 
dimensionality for the special multiports that optimally 
discriminate among the three quantum states.  For simplicity, we 
shall assume that the {\em a priori} probabilities are equal 
in all of our examples.

Our first case is a simple one; the overlaps of the three states 
will be assumed to be real and equal
\begin{equation}
\label{d1}
\langle \psi_1|\psi_2 \rangle
= \langle \psi_2|\psi_3 \rangle
=\langle \psi_3|\psi_1 \rangle
=s,
\end{equation}
where $0<s<1$. The constraint of equation\ (\ref{a8}) is,
in this case,
\begin{equation}
\label{d3}
q_1 q_2 q_3 - s^{2}\sum_{i} q_{i} +2s^{3}=0 ,
\end{equation} 
application of the the Lagrangian multiplier method implies
that $q_{1}=q_{2}=q_{3}$, and that
\begin{equation}
q_{i}^{3}-3s^{2}q_{i}+2s^{3}=0 .
\end{equation}
This equation has two solutions, $q_{i}=s, -2s$, of
which only $q_{i}=s$ is valid.  This solution is a
minimum and it implies that the optimal value of the 
total failure probability is $Q=s$.

Our next step is to find the failure vectors.  For any
$3\times 3$ positive matrix, $L$, we find that we can
express its matrix elements as $L_{ij}=\langle\phi_{i}
|\phi_{j}\rangle$ if
\begin{eqnarray}
\label{phi3}
|\phi_{1}\rangle & = & (\sqrt{L_{11}}, 0, 0) \nonumber \\
|\phi_{2}\rangle & = & \left( \frac{L_{12}}{\sqrt{L_{11}}},
\sqrt{\frac{\Delta_{12}}{L_{11}}}, 0\right) \nonumber \\
|\phi_{3}\rangle & = & \left( \frac{L_{13}}{\sqrt{L_{11}}},
\frac{L_{23}L_{11}-L_{12}^{\ast}L_{13}}
{\sqrt{L_{11}\Delta_{12}}},\frac{\Delta}{\Delta_{12}}\right) ,
\end{eqnarray}
where $\Delta_{12}=L_{11}L_{22}-|L_{12}|^{2}$ and $\Delta =
\det L$.  Applying this to the matrix $C$, with $q_{i}=
s$, $i=1,2,3$, we find that the three failure vectors are
identical, they all have magnitude $\sqrt{s}$ and point in
the same direction.  Therefore, our failure space, 
$\mathcal{A}$, is one dimensional, the full Hilbert space
$\mathcal{K}=\mathcal{H}\oplus\mathcal{A}$ is four
dimensional, and we will need an
eight port to accomplish our unitary transformation.

In order to find the necessary unitary transformation, we
must first specify our input states.  Let us choose our three 
states to be (in the full space, $\mathcal{K}=\mathcal{H}
\oplus\mathcal{A})$
\begin{eqnarray}
|\psi_{1}^{\mathcal{K}}\rangle_{in} =\left( 
\begin{array}{c}
\sqrt{\frac{2}{3}}\ \sqrt{1-s} \\
\frac{\sqrt{1+2s}}{\sqrt{3}} \\ 0 \\ 0\end{array}\right), &
|\psi_{2}^{\mathcal{K}}\rangle_{in} = \left(
\begin{array}{c} 
-\frac{\sqrt{1-s}}{\sqrt{6}} \\ \frac{\sqrt{1+2s}}{\sqrt{3}} \\
\frac{\sqrt{1-s}}{\sqrt{2}} \\ 0\end{array}\right) , \nonumber \\
|\psi_{3}^{\mathcal{K}}\rangle_{in} =\left( \begin{array}{c} 
-\frac{\sqrt{1-s}}{\sqrt{6}} \\ \frac{\sqrt{1+2\ s}}{\sqrt{3}}
\\ -\frac{\sqrt{1-s}}{\sqrt{2}} \\0 \end{array}\right) , &
\label{d6}
\end{eqnarray}
where $\psi_i$ are represented by single photon states.
One can verify that
$\langle \psi_1|\psi_2 \rangle
= \langle \psi_2|\psi_3 \rangle
=\langle \psi_3|\psi_1 \rangle
=s$.  The output states can be found from Eq. (\ref{b2}), and
are explicitly given by 
\begin{eqnarray}
\label{d7}
|\psi_{1}^{\mathcal{K}}\rangle_{out} = \left(
\begin{array}{c}
\sqrt{1-s} \\ 0 \\0 \\ \sqrt{s} \end{array}\right) ,&
|\psi_{2}^{\mathcal{K}}\rangle_{out} = \left(
\begin{array}{c}
0 \\ \sqrt{1-s} \\ 0 \\ \sqrt{s} \end{array}\right) ,
\nonumber \\
|\psi_{3}^{\mathcal{K}}\rangle_{out} = \left(
\begin{array}{c}
0 \\0 \\ \sqrt{1-s} \\ \sqrt{s} \end{array}\right) . & 
\end{eqnarray}
The unitary transformation, $U$, maps the input states onto
the output states, i.e.\ $|\psi_{i}^{\mathcal{K}}\rangle_{out}
=U|\psi_{i}^{\mathcal{K}}\rangle_{in}$, for $i=1,2,3$.  In
addition, it must map the vector that is orthogonal to the
three input vectors onto the vector that is orthogonal to 
the three output vectors, 
\begin{equation}
\frac{1}{\sqrt{2s+1}}\left(\begin{array}{c} \sqrt{s} \\
\sqrt{s} \\ \sqrt{s} \\ -\sqrt{1-s} \end{array}\right)
=U\left( \begin{array}{c} 0 \\ 0 \\ 0 \\ 1 \end{array}
\right) .
\end{equation}
The action of $U$ on these four vectors completely determines
it, and we find that it is given by the matrix $M(4)$, which is
\begin{equation}
\label{d9}
M(4) = \left(
\begin{array}{cccc}
\sqrt{\frac{2}{3}}
&\sqrt{\frac{1-s}{3(2s+1)}}
& 0 & \sqrt{\frac{s}{2s+1}} \\
-\frac{1}{\sqrt{6}}
& \sqrt{\frac{1-s}{3(2s+1)}}
&\frac{1}{\sqrt{2}}& \sqrt{\frac{s}{2s+1}} \\
-\frac{1}{\sqrt{6}}
& \sqrt{\frac{1-s}{3(2s+1)}}
&-\frac{1}{{\sqrt{2}}}& \sqrt{\frac{s}{2s+1}} \\
0&\sqrt{\frac{3s}{2s+1}} & 0 & -\sqrt{\frac{1-s}{2s+1}}
\end{array} \right) .
\end{equation} 
Using the method described in Sec. IV, $M(4)$ can be
factorized as
\begin{equation}
\label{d11}
M(4) = T_{1,2}\cdot T_{1,3}\cdot T_{2,3} \cdot T_{2,4} ,
\end{equation}
where the parameters that determine the matrixes $T_{pq}$
are given in table 1 (this example is referred to as case 1).  
Note that because these matrixes are
real, the complex conjugate, which appears in Eq. (\ref{mfac})
is unnecessary.

Now let us consider a more general case than the one we
have been studying so far.  We shall assume that two of
the overlaps are the same and the third is different, 
in particular that
\begin{eqnarray}
\langle\psi_{1}|\psi_{2}\rangle = & \langle\psi_{1}|
\psi_{3}\rangle & = s_{1} \nonumber \\
\langle\psi_{2}|\psi_{3}\rangle = & s_{2} & ,
\end{eqnarray}
where we shall assume, for simplicity, that $s_{1}$ and 
$s_{2}$ are real and between $0$ and $1$.  For a fixed value 
of $s_{1}$ there is a restriction on how large $s_{2}$ can be.
The largest the angle between $\psi_{2}$ and $\psi_{3}$ can
be is twice the angle between $\psi_{1}$ and $\psi_{2}$ (this
maximum is achieved when the vectors are coplanar).  This implies
that $s_{2}\ge 2s_{1}^{2}-1$.  Application of the Lagrange
multiplier method to the minimization of $Q^{\prime}$ gives
us $q_{2}=q_{3}$ and
\begin{equation}
q_{1}=\frac{q_{2}^{2}+s_{1}^{2}-s_{2}^{2}}{q_{2}} .
\end{equation}
Substituting these results into the constraint equation and
defining $y=q_{2}/s_{2}$ and $\beta = s_{1}/s_{2}$, 
we have
\begin{equation}
y^{4}-(2+\beta^{2} )y^{2}+2\beta^{2} y+1-\beta^{2} =0 .
\end{equation}
The roots of this equation are $y=1,\ -1\pm\beta$,
and two of them $1$ and $\beta-1$ yield valid
solutions, the latter if $\beta \ge 1$.  Substitution
of these results into $Q$ shows that if $\beta <2$, then
the solution $y=1$ gives the minimum and if $\beta \ge 2$,
then $y=\beta -1$ gives the minimum.  Summarizing, we
find that if $\beta < 2$, the minimum value of $Q$ is
$[(s_{1}^{2}/s_{2})+2s_{2}]/3$ and (solution $1$)
\begin{eqnarray}
q_{1} & = & \frac{s_{1}^{2}}{s_{2}} \nonumber \\
q_{2} & = & q_{3}= s_{2} ,
\end{eqnarray}
and if $\beta \ge 2$, then the minimum value of $Q$
is $2(2s_{1}-s_{2})/3$ and (solution $2$)
\begin{eqnarray}
q_{1} & = & 2s_{1} \nonumber \\
q_{2} & = & q_{3}=s_{1}-s_{2} .
\end{eqnarray}
Clearly, for these solutions to be valid, all of the
probabilities have to be between $0$ and $1$.

The next step is to find the failure vectors.  If solution
$1$ is the valid one, we find from Eq. (\ref{phi3}) that the 
failure space is one dimensional, and if 
$|u_{1}^{\mathcal{A}}\rangle$ is
the normalized basis vector for this space, then
\begin{eqnarray}
|\phi_{1}^{\mathcal{A}}\rangle & = & \frac{s_{1}}{\sqrt{s_{2}}}
|u_{1}^{\mathcal{A}}\rangle , \nonumber \\
|\phi_{2}^{\mathcal{A}}\rangle & = &
|\phi_{3}^{\mathcal{A}}\rangle =\sqrt{s_{2}}|u_{1}^{\mathcal{A}}
\rangle .
\end{eqnarray}
If solutions $2$ is the valid one, then the failure space is
two-dimensional.  If $|u_{j}^{\mathcal{A}}\rangle$ where $j=1,2$
is an orthonormal basis for this space we find that
\begin{eqnarray}
|\phi_{1}^{\mathcal{A}}\rangle & = & \sqrt{2s_{1}}
|u_{1}^{\mathcal{A}}\rangle , \nonumber \\
|\phi_{2}^{\mathcal{A}}\rangle & = & \sqrt{\frac{s_{1}}{2}}
|u_{1}^{\mathcal{A}}\rangle +\sqrt{\frac{s_{1}}{2}-s_{2}}
|u_{2}^{\mathcal{A}}\rangle , \nonumber \\
|\phi_{3}^{\mathcal{A}}\rangle & = & \sqrt{\frac{s_{1}}{2}}
|u_{1}^{\mathcal{A}}\rangle -\sqrt{\frac{s_{1}}{2}-s_{2}}
|u_{2}^{\mathcal{A}}\rangle .
\end{eqnarray}

Let us look at an example of each solution.  If
we choose our three states to be
$\psi_1=(1,0,0)$, $\psi_2=\frac{1}{\sqrt{3}}(1,1,1)$ and
$\psi_3=\frac{1}{\sqrt{3}}(1,1,-1)$, we find that $s_{1}=1/
\sqrt{3}$ and $s_{2}=1/3$, so that solution $1$ is valid.
The complete treatment of this case (case 2) is given in Table 2.
We see that we need an eight-port which can be built up 
by two $U(2)$ blocks.
Note that in order to achieve minimum failure probability
$Q$, we need to choose $q_1$ to be 1, which means that we
sacrifice the possibility of distinguishing state
$|\psi_1 \rangle$.

If we choose our states to be $\psi_1=(1,0,0)$, $\psi_2
=\frac{1}{3}(1,2,2)$ and $\psi_3=\frac{1}{3}(1,2,-2)$, then
we find that solution $2$ is valid with $s_{1}=1/3$ and
$s_{2}=1/9$.  In this case (case 3) we need a ten-port, and the 
complete results are given in Table 3.  Note that if the
procedure fails, it is still possible to gain some
information about the input state, because the failure
space is two-dimensional \cite{peres3}.  This is not possible 
if the failure space has only one dimension.  

One possibility
is to attach to the failure-space outputs (outputs 4 and 5) a
network that transforms states $|\phi_{2}^{\mathcal{A}}\rangle$
and $|\phi_{3}^{\mathcal{A}}\rangle$ into orthogonal states,
which it will do only with a certain probability \cite{bergou}.
In particular, we can construct a network that implements the
transformation
\begin{equation}
M(3)=\left( \begin{array}{ccc}
\frac{1}{\sqrt{6}} & \frac{1}{\sqrt{2}} & -\frac{1}{\sqrt{3}} \\
\frac{1}{\sqrt{6}} & -\frac{1}{\sqrt{2}} & -\frac{1}{\sqrt{3}} \\
\sqrt{\frac{2}{3}} & 0 & \frac{1}{\sqrt{3}} \end{array} \right) ,
\end{equation}
where the inputs to the first two ports of this network (we shall 
call these ports A and B) are the outputs of
ports 4 and 5 of the original network, and the input to the third 
port (port C) is the vacuum.  This network has been designed so that
if no photon is detected emerging from output C, then the input state 
$|\phi_{2}^{\mathcal{A}}\rangle$ will be transformed into a photon 
emerging from port A, and the input state $|\phi_{3}^{\mathcal{A}}
\rangle$ will be transformed into a photon emerging from port B.
If the input state is $|\phi_{1}^{\mathcal{A}}\rangle$ and no photon
is detected at output C, the probabilities of a photon emerging 
from either port A or port B are the same.  Therefore, if the photon
emerges from port A, we can conclude the input to the entire
network was either $\psi_{1}$ or $\psi_{2}$, and if it emerges from
port B, then the input was either $\psi_{1}$ or $\psi_{3}$.  
Summarizing, if one of the detectors in ports 1 through 3 clicks,
we know what the input state was.  If the detector in either ports
A or B clicks, then we gain partial information about the input
state; the number of possibilites has been reduced from three to
two.  If the detector in port C clicks, then we have gained no
information about the input state, and this happens with a 
probability of $1/9$ if the inputs were $\psi_{2}$ or $\psi_{3}$ 
and $4/9$ if the input was $\psi_{1}$.  The addition of the
second network to the failure outputs of the first significantly
improves the chances of gaining some information about the input
state.

\section{Conclusions}
We have shown that nonorthogonal quantum states, each realized 
as a photon split among several modes, can be conditionally
distinguished by means of a linear optical network.  For
three states we have given explicit networks, which give the
maximum success probabilities for several sets of states.  In
addition, it was shown that the addition of a second network to
the outputs corresponding to a failure of the initial network to
distinguish the states, can sometimes provide partial information 
about the input state.
We believe it should be possible to construct these networks in
the laboratory.

\section*{Acknowledgement}
This research was supported by the Office of Naval Research
(Grant Number: N00014-92J-1233), by the National Science
Foundation (Grant Number PHY-9970507 ) by the
Hungarian Science Research Fund (OTKA, Grant Number: T 030671)
and by a grant from PSC-CUNY as well as by a CUNY collaborative grant.
We would like to thank Gabriel Drobny for useful conversations.

\pagebreak

\begin{figure} 
\begin{center}
\epsfig{file=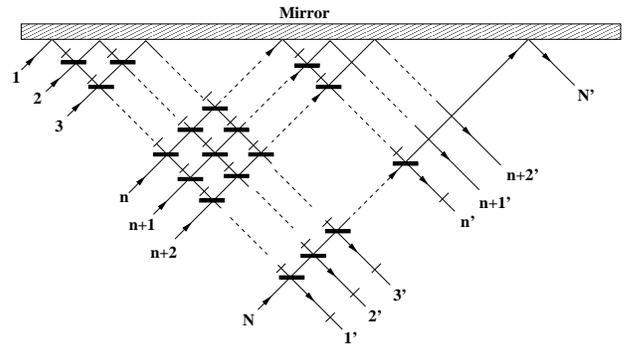, height=4.5cm}
\end{center}
\caption{The implementation of an optical multiport that performs the
unitary transformation $M(N)$ described in Eq. (\ref{c14}). 
The beams are straight lines, a suitable beam splitter
is at each crossing point of the first n diagonal lines, phase
shifters are at one input of each beam splitter and at the first n outputs.
Each diagonal line of the multiport reduces the dimension of $M(N)$ by
one.}
\label{fig1}
\end{figure}

\begin{table}
\caption{Arguments of $T_{p,q}$ for case 1. The range of arccos 
is the interval $[0,\pi ]$.}
%\[
\begin{tabular}{|>{$}c<{$}|>{$}c<{$}|>{$}c<{$}|}
\hline
& \omega &\phi 
\\ \hline
T_{1,2} & \arccos [-\frac{1}{\sqrt{5}}]
& 0 
\\ \hline
T_{1,3} & \arccos [-\frac{1}{\sqrt{6}}]
& 0
\\ \hline
T_{2,3} & -\arccos [- \sqrt{\frac{2}{5}}]
& 0 
\\ \hline
T_{2,4} & \arccos [\frac{\sqrt{3s}}{\sqrt{1+ 2s}}]
& 0 
\\ \hline
\end{tabular}
    \label{table1}
\end{table}

\pagebreak 

\begin{table}
    \label{table2}
\caption{Summary of parameters and arguments of $T_{p,q}$ for
case 2.}
%\setlength{\extrarowheight}{2pt}
%\[
\begin{tabular}{|c|>{$}c<{$}|} \hline
&\text{case 2}
\\  \hline
Input states
&
\begin{tabular}{>{$}c<{$}>{$}c<{$}>{$}c<{$}}
\psi_1 & \psi_2 &\psi_3
\\
\left(\begin{array}{c}1 \\ 0 \\ 0 \\ 0  \end{array}\right)
&
\left(\begin{array}{c} \frac{1}{\sqrt{3}} \\ \frac{1}{\sqrt{3}}
    \\ \frac{1}{\sqrt{3}} \\ 0 \end{array} \right)
&
\left(\begin{array}{c}\frac{1}{\sqrt{3}} \\ \frac{1}{\sqrt{3}}
    \\-\frac{1}{\sqrt{3}} \\ 0\end{array} \right)
\end{tabular}

\\  \hline
\begin{tabular}{c}
Optimal failure \\ probability
\end{tabular}
& 
\begin{tabular}{>{$}c<{$}}
q_1 =1 \\ q_2 =\frac{1}{3} \\  q_3 =\frac{1}{3}
\end{tabular}

\\  \hline
Output states
&
\begin{tabular}{>{$}c<{$}>{$}c<{$}>{$}c<{$}}
\psi_1 & \psi_2 &\psi_3
\\
\left(\begin{array}{c} 0 \\ 0 \\ 0 \\ 1 \end{array} \right)
&
\left(\begin{array}{c} 0 \\\sqrt{\frac{2}{3}}\\ 0 \\ \frac{1}{\sqrt{3}}
\end{array} \right)
&
\left(\begin{array}{c} 0 \\ 0 \\ \sqrt{\frac{2}{3}} \\ \frac{1}{\sqrt{3}}
\end{array} \right)
\end{tabular}

\\  \hline
$M=$
&
\left(\begin{array}{cccc}
0&0&0&1
\\0&\frac{1}{\sqrt{2}}&\frac{1}{\sqrt{2}}&0
\\0&\frac{1}{\sqrt{2}}&-\frac{1}{\sqrt{2}}&0
\\1&0&0&0
\end{array} \right)

\\ \hline
\begin{tabular}{c}
Factorization\\of $M$
\end{tabular}
&
M =T_{1,4} \cdot T_{2,3}

\\ \hline

\begin{tabular}{c}
Arguments \\of $T_{p,q}$
\end{tabular}
&
\begin{tabular}{>{$}c<{$}}
T_{1,4}:{\omega=0, \phi=0}
\\ T_{2,3}:{\omega=\frac{\pi}{4}, \phi=0}
\end{tabular}

\\  \hline
\end{tabular}            %\]
\end{table}

\newpage 
\begin{table}
\caption{Summary of parameters and arguments of $T_{p,q}$ for
case 3.}
\setlength{\extrarowheight}{2pt}
\[
\begin{tabular}{|c|>{$}c<{$}|} \hline
&\text{case 3}
\\  \hline
Input states
&
\begin{tabular}{>{$}c<{$}>{$}c<{$}>{$}c<{$}}
\psi_1 & \psi_2 &\psi_3
\\
\left(\begin{array}{c}1 \\0 \\0 \\0 \\0   \end{array}\right)
&
\left(\begin{array}{c} \frac{1}{3} \\\frac{2}{3} \\
\frac{2}{3} \\
0 \\
0  \end{array}\right)
&
\left(\begin{array}{c}\frac{1}{3} \\\frac{2}{3} \\-\frac{2}{3} \\0 \\0 \end{array}\right)
\end{tabular}

\\  \hline
\begin{tabular}{c}
Optimal failure \\ probability
\end{tabular}
& 
\begin{tabular}{>{$}c<{$}}
q_1 =\frac{2}{3} \\ q_2=\frac{2}{9} \\  q_3=\frac{2}{9}
\end{tabular}

\\  \hline
Output states
&
\begin{tabular}{>{$}c<{$}>{$}c<{$}>{$}c<{$}}
\psi_1 & \psi_2 &\psi_3
\\
\left(\begin{array}{c}	\frac{1}{{\sqrt{3}}} \\
	0 \\
	0 \\
	{\sqrt{\frac{2}{3}}} \\
	0   \end{array}\right)
&
\left(\begin{array}{c} 0 \\ \frac{{\sqrt{7}}}{3} \\0 \\
\frac{1}{{\sqrt{6}}} \\
\frac{1}{3\ {\sqrt{2}}}   \end{array}\right)
&
\left(\begin{array}{c}0 \\
	0 \\
	\frac{{\sqrt{7}}}{3} \\
	\frac{1}{{\sqrt{6}}} \\
	-\frac{1}{3\ {\sqrt{2}}}  \end{array}\right)
\end{tabular}

\\  \hline
$M$
&
\left(\begin{array}{ccccc}
\frac{1}{{\sqrt{3}}}&-\frac{1}{2{\sqrt{3}}}&0&-{\sqrt{\frac{7}{12}}}&0 \\
0&\frac{{\sqrt{7}}}{4}&\frac{{\sqrt{7}}}{4}&-\frac{1}{4}&\frac{1}{4} \\
0&\frac{{\sqrt{7}}}{4}&-\frac{{\sqrt{7}}}{4}&-\frac{1}{4}&-\frac{1}{4} \\
{\sqrt{\frac{2}{3}}}&\frac{1}{2\ {\sqrt{6}}}&0&{\sqrt{\frac{7}{24}}}&0 \\
0&0&\frac{1}{2\ {\sqrt{2}}}&0&-{\sqrt{\frac{7}{8}}} 
\end{array}\right)

\\ \hline
\begin{tabular}{c}
Factorization\\of $M$ 
\end{tabular}
&
M =T_{1,4} \cdot T_{2,3} \cdot T_{2,4}\cdot T_{3,5}

\\ \hline

\begin{tabular}{c}
Arguments \\of $T_{p,q}$
\end{tabular}
&
\begin{tabular}{>{$}l<{$}}
T_{1,4}:{\omega=\arccos \big[{\sqrt{\frac{2}{3}}}\big]
, \phi=0}
\\ T_{2,3}:{\omega=\frac{\pi }{4}, \phi=0}
\\ T_{2,4}:{\omega=\arccos \big[-\frac{1}{2\ {\sqrt{2}}}\big], \phi=0}
\\ T_{3,5}:{\arccos \big[\frac{1}{2\ {\sqrt{2}}}\big], \phi=0}
\end{tabular}

\\  \hline
\end{tabular}
		    \]
\end{table}

\end{document}